\documentclass[twocolumn,showpacs,preprintnumbers,amsmath,amssymb]{revtex4}

\usepackage{graphicx}
\usepackage{dcolumn}
\usepackage{bm}

\begin{document}


\title{Universal Extremal Statistics in a Freely Expanding Jepsen Gas}

\author{Ioana Bena}
\email{Ioana.Bena@physics.unige.ch}
\affiliation{
D\'epartement de Physique Th\'eorique, Universit\'e de Gen\`eve, CH-1211 Gen\`eve 4, Switzerland}
\author{Satya N. Majumdar}
 \email{satya.majumdar@u-psud.fr}
\affiliation{Laboratoire de Physique Th\'{e}orique et Mod\`{e}les
Statistiques, Universit\'{e} Paris-Sud, B\^{a}timent 100, 91405
Orsay Cedex\\ France}

\date{\today}

\begin{abstract}

We study the extremal dynamics emerging 
in an out-of-equilibrium one-dimensional 
Jepsen gas of $(N+1)$ hard-point particles. 
The particles undergo binary 
elastic collisions, but move ballistically 
in-between collisions. The gas is initally uniformly 
distributed in a box $[-L,0]$ with the 
``leader" (or the rightmost particle) at $X=0$, and a 
random positive velocity, independently drawn from 
a distribution $\phi(V)$, is assigned to each particle. 
The gas expands freely at subsequent times. We compute
analytically the distribution of the leader's velocity 
at time $t$, and also
the mean and the variance of the number of collisions 
that are undergone by the leader
up to time $t$. We show that in the thermodynamic limit 
and at fixed time $t\gg 1$ (the so-called ``growing regime"), 
when interactions are strongly manifest, 
the velocity distribution exhibits universal 
scaling behavior of only three possible
varieties, depending on the tail of $\phi(V)$. 
The associated scaling functions
are novel and different from the usual 
extreme-value distributions of uncorrelated
random variables. In this growing
regime the mean and the variance 
of the number of collisions of the leader up to time 
$t$ increase logarithmically 
with $t$, with universal prefactors that are computed exactly. 
The implications of our results in the context
of biological evolution modeling are pointed out. 
\end{abstract}


\maketitle

\section{Introduction}
\label{introduction}

The study of the statistics of the maximum or the minimum 
of a set of random variables,
generally referred to as the extreme-value statistics (EVS), 
is important in diverse areas including  
disordered systems such as spin-glasses~\cite{bouchaud}
and directed polymers~\cite{Johansson,mk1,satya3}, 
turbulent flows~\cite{pinton},
sorting and search problems in computer science~\cite{satya2,satya4,mk2}, 
fluctuating interfaces~\cite{racz1,racz2,bolech,guclu,satya5,satya6,racz3}, 
granular matter~\cite{anna},
growing networks~\cite{Redner1}, and models of 
biological evolution~\cite{GE,krug,jain,satya1}.
The theory of EVS is simple and well understood~\cite{gumbel,coles} 
when the $N$ random variables
$V_1$, $V_2$, $\dots$, $V_N$ are statistically {\em uncorrelated} and each of them
is drawn from the same common parent distribution $\phi(V)$. Their maximum
$V_{\rm max}= {\rm max}(V_1,V_2,\dots,V_N)$ is a random variable
whose probability density $P(V,N)dV = 
{\rm Prob}[V\leqslant V_{\rm max}\leqslant V+dV,N]$ is known to 
have a scaling form for large $N$,
\begin{equation}
P(V,N)\approx \frac{1}{{\tilde 
b}_i(N)}\;G_i\left(\frac{V-{\tilde a}_i(N)}{{\tilde
b}_i(N)}\right)\,,\;\;i=I,II,III\,.
\label{pden1}
\end{equation}
Here the subscript $i$ 
refers to the three types of tails of the parent distribution $\phi(V)$, 
namely:\\ 
\noindent(I) a tail decaying faster than a power law, 
such as $\phi(V)\propto V^{\beta}\exp\left(-V^{\delta}\right)$ when
$V\to \infty$, with $\delta>0$;\\
\noindent (II) a power-law tail, such as $\phi(V)\propto V^{-\beta}$ 
as $V\to \infty$, where $\beta>2$;\\
and\\
\noindent(III) a bounded distribution, such as $\phi(V)\propto (V_c-V)^{\beta}$
when $V\to V_c^{-}$, with $\beta>0$.\\
The functions ${\tilde a}_i(N)$-s and ${\tilde b}_i(N)$-s are 
non-universal scale factors that depend on 
the details of $\phi(V)$. However, the scaling functions $G_i(z)$-s
are universal~\cite{gumbel,coles}, in the sense that they are only 
of three possible varieties,
depending on the three classes I, II, and III, but are otherwise 
independent of the details of $\phi(V)$.
These universal functions are known, respectively, as 
(I) Fisher-Tippett-Gumbel, (II) Fr\'echet, and (III) Weibull 
distributions.

In contrast to the above case of independent identically-distributed
(i.i.d.) stochastic variables, the EVS is much less understood when 
there are {\em correlations} or {\em interactions} between the random variables. 
In the presence of static interactions,
exact results for the EVS are known only in few cases, such as for fluctuating
$(1+1)$-dimensional interfaces in their steady states~\cite{satya5,satya6,racz3}
and for a class of directed polymer problems~\cite{Johansson,mk1,satya3}.
In this paper, we present exact asymptotic results for the EVS in an
interacting particle system, for which the interactions
between the random variables are manifest {\em dynamically}.
Our system is a one-dimensional gas consisting 
of $(N+1)$ identical hard-point particles that undergo
binary elastic collisions. 
At these instantaneous collisions the particles thus merely 
exchange their velocities, while in-between collisions they 
move freely. Due to the simplicity of the dynamics, 
this so-called  {\em Jepsen gas}
often admits analytical treatments for various externally imposed
constraints, and hence has a rather rich history, see 
e.g.~\cite{frisch,teramoto,jepsen,lebowitz1,lebowitz2,mckean,keyes}.
It has also proved very useful in the study of 
a class of out-of-equilibrium problems like the 
evolution of the ``adiabatic piston"~\cite{jarek1,jarek2,bala1,bala2},  
the Jarzynksi theorem~\cite{ioana}, and 
a quasispecies biological evolution model~\cite{krug,jain,satya1}. 
Moreover, the Jepsen gas also turns out to have important applications in
spin transport processes in the one-dimensional nonlinear 
$\sigma$-model~\cite{sachdev1,sachdev2}. 

\begin{figure}
\begin{center}
\includegraphics[width=0.88\columnwidth]{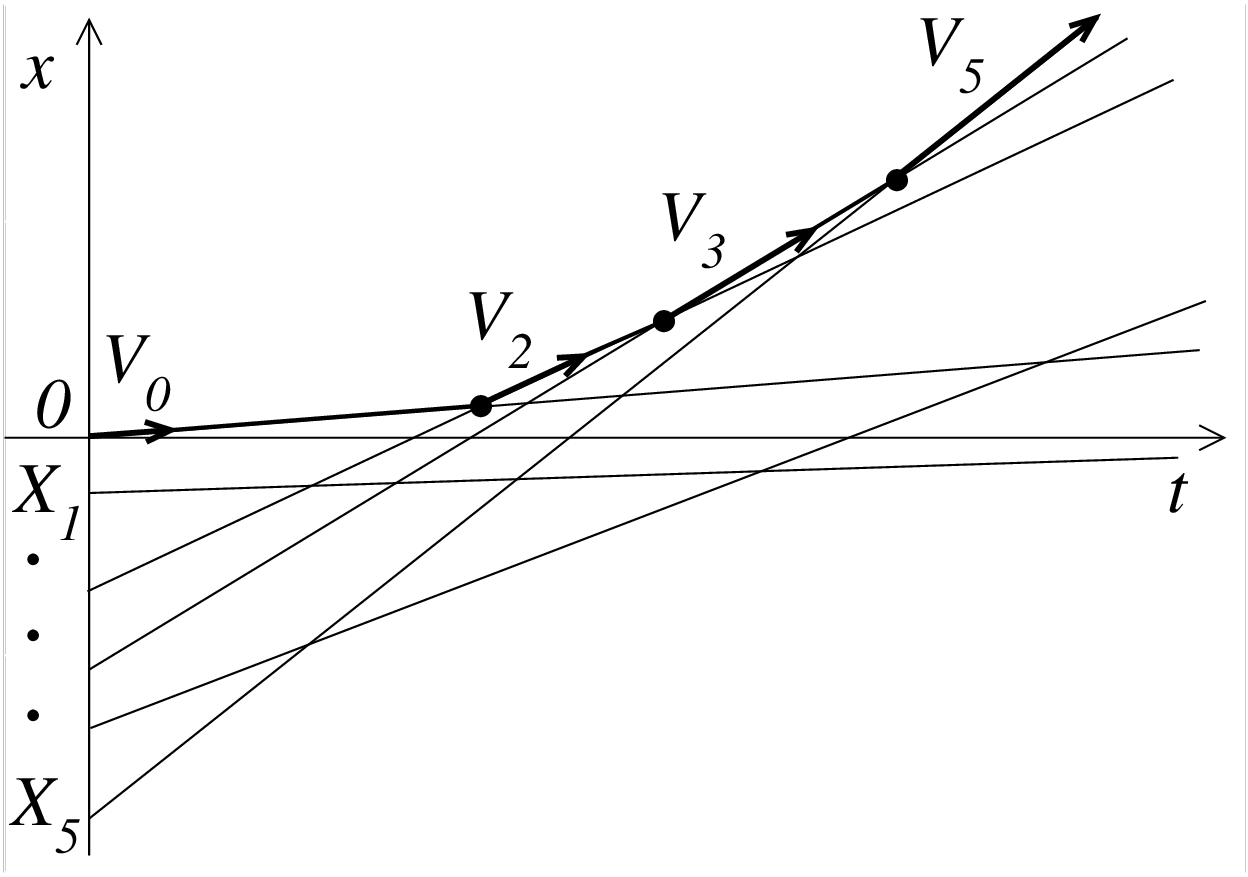}
\caption{\label{figure1} A realization of the
trajectories for $N=5$ particles
plus the leader. 
The thick line indicates
the trajectory of the leader, and the labelled arrows refer to 
the successive values of its velocity. The points where
the leader trajectory gets modified through collisions
are indicated by the big black dots.
The thin lines correspond to the trajectories of the other particles.}
\end{center}
\end{figure}

We start from an initial condition at $t=0$ where 
the Jepsen gas of $(N+1)$ particles
occupies the interval $[-L,0]$ on the real $x$ axis. 
The extreme-right ``zero"-th 
particle, that we shall conventionally 
call hereafter the {\em leader}, is  initially located 
at $X_0=0$.
The coordinates $X_i$, $i=1,...,N$ of the $N$ particles 
to the left of the leader are uniformly distributed 
in the interval $-L \leqslant x < X_0=0$. Thus the gas
has a uniform initial density $n_0=N/L$. 
The initial velocities $V_i$, $i=0,...,N$ of the leader 
and of the other $N$ particles are independent random variables, 
identically distributed according to the {\em parent distribution} 
$\phi(V)$. 
For simplicity, we restrict ourselves here to the case of
positive velocities, i.e., such that $\phi(V)=0$ for $V<0$.
No boundaries affect the dynamics of the system,
which is simply a {\em free expansion}. 
As the system evolves in time,
the particles collide elastically and exchange their velocities. 
For a given initial condition, the system up to time $t$ is fully
described by the set of trajectories $\{X_i+V_i t,\, i=0,1,\dots, N\}$.
Each of the particles travels along such a trajectory till it collides
with another particle, and each collision changes its trajectory.
In particular, the velocity of the leader increases
whenever it collides with a particle 
with higher velocity coming
from its left, see Fig. \ref{figure1}. 
For a fixed particle number $N$, it is obvious that
if one waits for a long enough time, then the leader
will acquire the largest velocity of the initial set 
$\{V_i,\,i=0,1,\dots N\}$.
Once this happens, the trajectory of the
leader remains unchanged for all subsequent times.

In the context of biological evolution of quasispecies, this model was first
introduced and studied in Ref.~\cite{krug}. 
The population $n_i(t)$ of the $i$-th genotype or species
increases exponentially with time, $n_i(t)=n_i(0)\, \exp(V_i t)$ where
the effective rate of reproduction $V_i\geqslant 0$ defines its ``fitness".
The logarithmic variable $X_i(t)= \ln[n_i(t)]= X_i(0)+ V_i\, t$ can then
be interpreted as the $i$-th trajectory of a Jepsen gas. At $t=0$ the
rightmost particle with $X_0=0$ and velocity $V_0$ is the leader/most fitted
genotype; however, if $V_0$ is not the maximal of the initial velocities, 
then it will be overtaken 
successively by faster/better fitted genotypes. 
At each of these overtaking events
the velocity of the leader changes instantaneously by a finite amount, 
i.e., the fitness of the leading genotype increases discontinuously.
These overtaking events represent thus the so-called punctuation events 
in the general context of evolution~\cite{GE}. 
The important observable here is the number of leading genotype 
changes up to time $t$, i.e., the number of collisions that the 
leader undergoes up to time $t$. In particular, the total 
number of punctuation events till the emergence of the 
eventual ``absolute" leader genotype exhibits universal dependencies 
on the system size $N$, that were investigated numerically 
in Refs.~\cite{krug,jain}
and recently analytically in Ref.~\cite{satya1}.

The goal of this paper is to compute analytically 
two physical quantities of principal interest, namely:\\ 
(A) the probability density of the velocity of the leader 
$P(V,t,N)$ at any time $t$, and for a fixed number $(N+1)\gg 1$ of particles, 
and \\
(B) the mean and the variance of the number of collisions 
undergone by the leader (i.e., the number of leader's trajectory
changes) up to a finite time $t$. 

It is clear that in this system,
as long as $N \gg 1$ is finite, there is a natural time scale 
$t^*(N) \gg 1$ 
that denotes the time taken by the final trajectory of the leader
to emerge. 
For $t>t^*(N)$, the leader velocity does not change anymore,
and as such the leader follows on the trajectory corresponding to 
the maximum of the initial velocities. Therefore, there
are obviously two temporal regimes separated by the crossover 
time scale $t^*(N)$, namely:

\vspace{0.2cm}

\noindent{\bf The stationary regime ($t>t^*(N)\gg 1$):} 
In this regime, the leader's trajectory remains unchanged for 
all subsequent times.  
The probability density of the leader velocity $P(V,N,t)$ becomes 
thus time-independent for $t>t^*(N)\gg 1$,
and it is given by the probability density $P(V,N)$ 
of the maximum of the initial velocities.
Since the initial velocities are i.i.d., $P(V,N)$
satisfies the scaling form in Eq.~(\ref{pden1}), 
where the scaling function $G_i(z)$
has one of the three universal forms 
(Fisher-Tippett-Gumbel, Fr\'echet, or Weibull), depending on the 
tail of $\phi(V)$. 
Thus, one obtains this regime by taking $t\to \infty$ limit, 
but keeping $N\gg 1$ fixed.
The interactions between particles, 
which are manifest only dynamically, become
completely irrelevant in this regime for the 
velocity distribution of the leader. 
 
\vspace{0.2cm}
\noindent{\bf The growing regime ($1 \ll t \ll t^*(N)$):} 
In this regime the interactions
play an important role, and thus the velocity distribution 
of the leader $P(V,t,N)$ is nontrivial. Since $t \ll t^*(N)$,
the finitness of the system does not affect the leader dynamics; 
as such, one can study this regime by
fixing $t$ and taking the thermodynamic limit, 
i.e., $N\to \infty$ limit (at fixed $n_0=N/L$). One can show that
the leader velocity distribution approaches 
an $N$-independent form $P(V,t)$. In addition,
for $t \gg 1$ (still in the thermodynamic $N\to \infty$ limit), 
$P(V,t)$ has a scaling form
\begin{equation}
P(V,t) \approx
\frac{1}{b_i(t)}\;F_{i}\left(\frac{V-a_i(t)}{b_i(t)}\right)\,,\;\;i=I,II,III\,,
\label{scaling}
\end{equation}
where the index $i=I,II, III$ refers to the three types of 
tails of the parent distribution 
$\phi(V)$ as mentioned before. 
The functions $a_i(t)$-s and $b_i(t)$-s are non-universal 
scale factors that depend on the details of $\phi(V)$, 
but the scaling functions
$F_i(z)$-s are universal and are of one of three types
I, II, and III, but are otherwise independent of the
details of $\phi(V)$. 
Moreover, the scaling functions $F_i(z)$-s are different
from the scaling functions $G_i(z)$-s in Eq.~(\ref{pden1}) 
that characterize the EVS of i.i.d random 
variables. The exact forms of the scaling functions 
$F_i(z)$-s are detailed in the next section.
 
We have also computed analytically the mean 
and the variance of the number of
collisions $n_c(t,N)$ undergone by the 
leader up to time $t$ in a system
with $(N+1)\gg 1$ particles. In the stationary 
regime $t>t^*(N)\gg 1$, it is
already known that the statistics 
of $n_c(t,N)$ become time-independent
and presents universal $N$-dependence~\cite{krug,jain,satya1}. 
More precisely,
$\langle n_c(N)\rangle \approx \xi_i \ln(N)\,,\quad i=I,II,III$ for large $N$,
where the prefactor $\xi_i$ is universal and its value is specific to
each of the three classes $i=I,II, III$.
This result was first conjectured in Ref.~\cite{krug}
and was later on proved analytically in Ref.~\cite{satya1}. 

In this paper, we compute
$\langle n_c(t,N)\rangle $ in the growing regime 
($1 \ll t \ll t^*(N)$) and show that
the corresponding mean number of collisions 
is independent of $N$ for large $N$, and grows
universally with time as 
$\langle n_c(t)\rangle \approx \gamma_i \ln(n_0 t)\,,\quad i=I,II,III$,
where $n_0$ is the initial density, 
and the prefactor $\gamma_i$ is universal
and dependent on the type of tail of $\phi(V)$.
Similar universal results are also derived 
for the variance of $n_c(t,N)$ in the
growing regime (see Secs.~II and IV for details).

The paper is organized as follows. First, for ready reference, 
we present in Sec.~II a summary of our main results. 
In Sec.~\ref{pdfsection}, we compute the velocity distribution
of the leader and provide exact asymptotic results.
Section~\ref{collision} discusses the
statistics of the number of colllisions undergone by the leader 
up to time $t$.
In Sec.~\ref{example} we provide, for illustration, the full  
analytical results for a particular choice of $\phi(V)$.
Finally, we conclude with a summary and outlook in Sec.~\ref{end}.
To facilitate the reading, most of the lengthy 
calculations are relegated to the
Appendices~A--D.

\section{Summary of the main results}
\label{summary}

In this Section, we summarize our main results for\\ 
(A) the velocity distribution
of the leader, and \\
(B) the mean and the variance 
of the number of collisions undergone
by the leader up to time $t$. 

\subsection{The asymptotic velocity distribution of the leader}
\label{long}

As mentioned in the Introduction, 
the asymptotic velocity distribution of the leader, both in the 
stationary as well as in the growing regime, 
is universal, in the sense that it depends only
on the tail of the parent distribution $\phi(V)$. 
As such, three different universality classes
emerge. We summarize below these universal behaviors 
for these three classes.
\quad\\
\noindent{\bf Class I: Tail decaying faster than a 
power-law as $V\to \infty$}, 
\begin{equation}
\phi^{\rm{tail}}(V)={\cal A}V^{\beta}e^{-V^{\delta}}
\label{class1}
\end{equation}
with $\delta >0$ and ${\cal A}$ a constant related to the details 
(in particular, the normalization factor)
of $\phi(V)$. In this case, the crossover time 
scales with $N$ as $t^*(N)\propto N$
for large $N$.
 
\vspace{0.2cm}

\noindent {\em The stationary regime ($t \gg t^*(N) \gg 1$):} 
the probability density $P(V,N)$ of the leader
becomes time-independent and is given by the EVS of 
the i.i.d. initial velocities. 
In the limit of large $N$
(see Sec.~IVB for details),  $P(V,N)$ 
satisfies the
scaling form in Eq. (\ref{pden1}) with 
\begin{eqnarray}
{\tilde a}_I(N) &\approx& \left[\ln \left(\frac{N {\cal A}}{\delta}\right)\right]^{1/\delta}\,,\nonumber\\
{\tilde b}_I(N) &\approx& \frac{1}{\delta}\left[\ln \left(\frac{N {\cal A}}
{\delta}\right)\right]^{(1-\delta)/\delta}\,\,,
\label{sc1}
\end{eqnarray}
and the scaling function is equal to 
the Fisher-Tippett-Gumbel probability density function (p.d.f.),
\begin{equation}
G_I(z)= \exp\left[-z-\exp(-z)\right]\,,\quad-\infty < z < \infty.
\label{gumbel1}
\end{equation}

\vspace{0.2cm}
 
\begin{figure}
\begin{center}
\includegraphics[width=0.88\columnwidth]{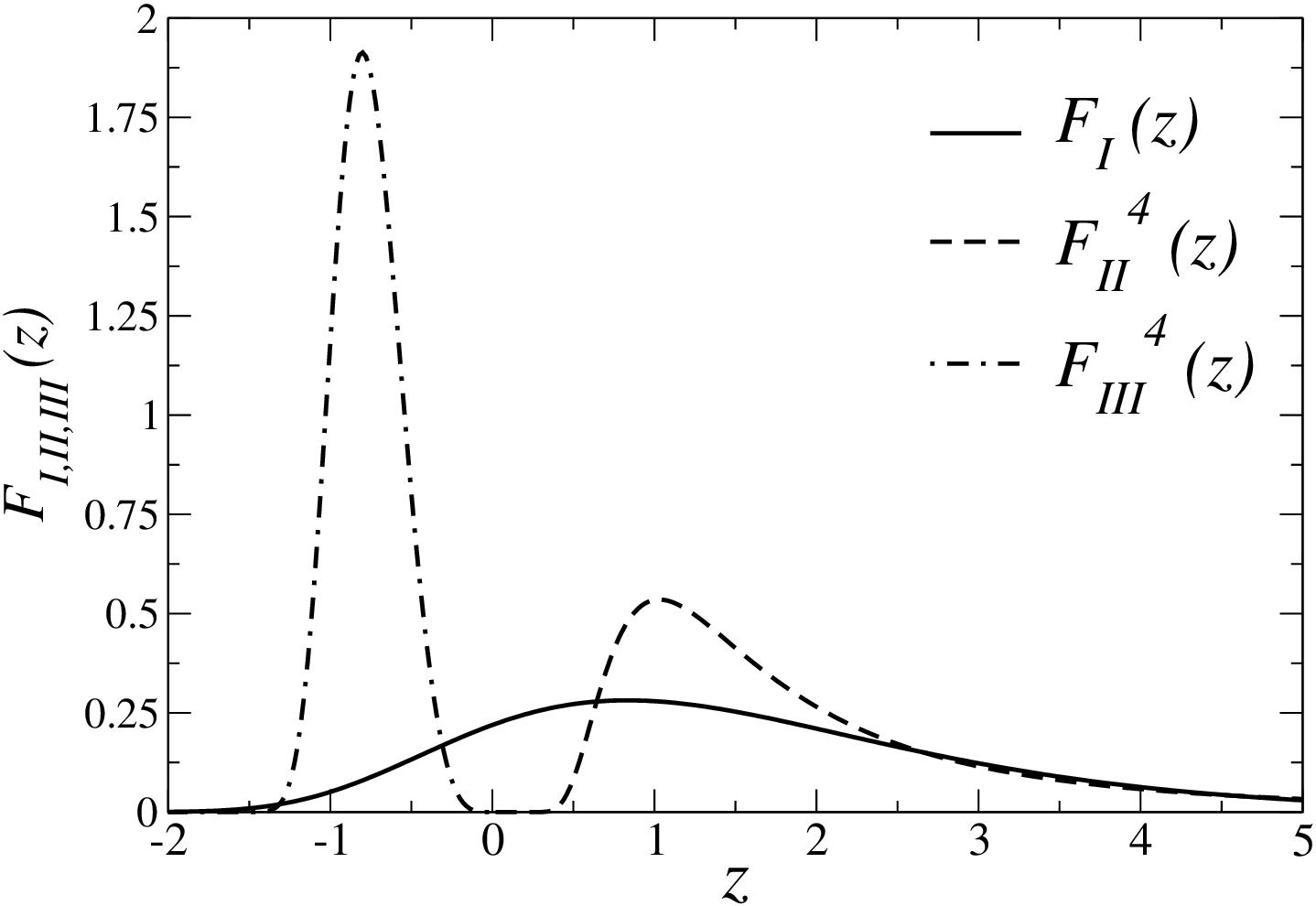}
\caption{\label{figure2} The universal functions $F_I(z)$ 
(with continuous line),  
$F_{II}^4(z)$ (dashed line), 
and $F_{III}^4(z)$ (dashed-dotted line).
The superscript ``4" represents  the value of the 
parameter $\beta$.}
\end{center}
\end{figure}

\noindent {\em The growing regime ($1 \ll t \ll t^*(N)$):} 
the velocity distribution
$P(V,t)$ of the leader becomes independent of $N$ as $N\to \infty$ 
(with $t$ fixed),
and has the scaling form described in Eq.~(\ref{scaling}) with
\begin{eqnarray}
a_I(t) &\approx& \left[\ln\left(\frac{n_0t{\cal
A}}{\delta^2}\right)\right]^{1/\delta}\,,\nonumber\\
b_I(t) &\approx& \frac{1}{\delta} 
\left[\ln\left(\frac{n_0t{\cal A}}{\delta^2}\right)\right]^{(1-\delta)/\delta}\,,
\label{par1}
\end{eqnarray}
and with the universal scaling function     
\begin{eqnarray}
F_I(z)=e^{-z}\int_{-\infty}^z dU \displaystyle e^{-e^{-U}}&=&-e^{-z}{\rm{Ei}}(-e^{-z})\,,
\nonumber\\
&&-\infty < z < \infty \,,
\label{f1}
\end{eqnarray}
where ${\rm Ei}(z)$ is the exponential-integral function~\cite{grad}. 
The profile of $F_I(z)$ is represented in Fig.~\ref{figure2},
and its asymptotics are
\begin{equation}
F_I(z)\approx \left\{
\begin{array}{ll}
(z-C)e^{-z}& \rm{for}\quad z \rightarrow \infty\,,\\
& \\
\displaystyle \exp({-e^{-z}}) &\rm{for} \quad z \rightarrow -\infty\,,
\end{array}
\right.
\end{equation}
where $C=0.577215...$ is Euler's constant.
Correspondingly, the mean asymptotic velocity of the leader increases 
with time $t$ as
\begin{equation}
\langle V (t)\rangle \approx \left[\ln (n_0t)\right]^{1/\delta}\,,
\label{vel1}
\end{equation}
and naturally this is also independent on other details of $\phi(V)$.

\quad\\
\noindent{\bf Class II: Power-law-decaying distributions $\phi(V)$ as $V\to \infty$,}
\begin{equation}
\phi^{\rm{tail}}(V)={\cal B} V^{-\beta}
\label{class2}
\end{equation}
where $\beta>2$ in order to ensure the finitness of the first moment of
$\phi(V)$,
and ${\cal B}$ is a constant.
In this case the crossover time $t^*(N) \approx N^{(\beta-2)/(\beta-1)}$ 
for large $N$, and the leader velocity distribution has the 
following asymptotic universal form in the two regimes: 

\vspace{0.2cm}

\noindent{\em Stationary regime:} the time-independent $P(V,N)$  
satisfies the scaling form in Eq.~(\ref{pden1}), with
\begin{eqnarray}
{\tilde a}_{II}(N) &=& 0\,,\nonumber\\
{\tilde b}_{II}(N) &\approx& {\left(\frac{{\cal B} N}{\beta-1}\right)}^{1/(\beta-1)}, 
\label{sc2}
\end{eqnarray}
and the scaling function is the Fr\'echet p.d.f.,
\begin{equation}
G_{II}(z)=\left\{
\begin{array}{ll}
\displaystyle\frac{(\beta-1)}{z^{\beta}}\,
\exp\left(-\displaystyle\frac{1}{z^{\beta-1}}\right) &\rm{for}\quad z \geqslant 0\,,\\
 & \\
\displaystyle 0 &\rm{for} \quad z < 0\,.
\end{array}
\right.
\label{frechet1}
\end{equation}

\vspace{0.2cm}

\noindent {\em Growing regime:} the leader velocity distribution $P(V,t)$ 
becomes independent of $N$ as $N\to \infty$, 
and satisfies Eq.~(\ref{scaling}) with
\begin{equation}
a_{II}(t)=0\,,\;
b_{II}(t) \approx \displaystyle\left[\frac{n_0t\,{\cal B}}{(\beta-1)(\beta-2)}\right]^{1/(\beta-2)}\,.
\label{par2}
\end{equation}
The universal scaling function $F_{II}(z)$ depends only on the
parameter $\beta$ (but is otherwise independent of the details of $\phi(V)$):
\begin{equation}
F_{II}^{\beta}(z)=(\beta-1)z^{-\beta}\int_{z^{2-\beta}}^\infty dU 
\;U^{-(\beta-1)/(\beta-2)}\,e^{-U}\,,
\end{equation}
that can also be written as
\begin{equation}
F_{II}^{\beta}(z)=(\beta-1)(\beta-2)z^{-\beta}\int_0^z dU\,e^{-U^{2-\beta}}\,,
\label{f2}
\end{equation} 
and is defined for $z\in [0,\infty)$. 
For $z < 0$, $F_{II}^{\beta}(z)=0$ trivially.
Its limiting behavior is given by 
\begin{equation}
F_{II}^{\beta}(z)\approx \left\{
\begin{array}{ll}
\displaystyle\frac{(\beta-1)(\beta-2)}{z^{\beta-1}}& \rm{for}\quad z \rightarrow \infty\,,\\
 & \\
\displaystyle\frac{\beta-1}{z}\exp(-z^{2-\beta}) &\rm{for} \quad z \rightarrow
0^+\,.
\end{array}
\right.
\end{equation}
The mean velocity of the leader increases asymptotically as
a power-law,
\begin{eqnarray}
\langle V (t)\rangle &\approx& \frac{\beta-1}{\beta-2}\;
\Gamma\left(\frac{\beta-3}{\beta-2}\right)\;b_{II}(t) \nonumber\\
&\approx& \frac{\beta-1}{\beta-2} \left[\frac{{\cal B}}
{(\beta-1)(\beta-2)}\right]^{1/(\beta-2)}
\Gamma\left(\frac{\beta-3}{\beta-2}\right)\nonumber\\
&\times&\;(n_0t)^{1/(\beta-2)}\,.\nonumber\\
\label{vel2}
\end{eqnarray}
The above expression is finite provided that $\beta>3$. Thus,
although  $P(V,t)$ exists for 
all $\beta>2$, its first moment is infinite for 
$2 < \beta \leqslant 3$.

\quad\\
\noindent{\bf Class III: Distribution $\phi(V)$
with a finite maximum velocity $V_c$,}
\begin{equation}
\phi^{\rm{tail}}(V)={\cal C}\,(V_c-V)^{\beta} \quad \mbox{for}\;\;
V\lesssim V_c\,,
\label{class3}
\end{equation}
with $\beta >0$. 
In this case, $t^*(N)\approx N^{(\beta+2)/(\beta+1)}$ for large $N$.

\vspace{0.2cm}

\noindent{\em Stationary regime:}  $P(V,N)$ satisfies the scaling
form of Eq.~(\ref{pden1}) with
\begin{eqnarray}
{\tilde a}_{III}(N) &=& V_c\,,\nonumber\\
{\tilde b}_{III}(N) &\approx& {\left(\frac{{\cal C} N}{\beta+1}\right)}^{-1/(\beta+1)}\,,
\label{sc3}
\end{eqnarray}
and the scaling function is the Weibull p.d.f.,
\begin{equation}
G_{III}(z)=\left\{
\begin{array}{ll}
 0 & \rm{for}\quad z \geqslant 0\,,\\
 & \\
(\beta+1)\,|z|^{\beta}\, \exp\left[-|z|^{\beta+1}\right] & \rm{for} \quad z <
0\,.
\end{array}
\right.
\label{wiebull1}
\end{equation}

\vspace{0.2cm}

\noindent {\em Growing regime:} In this regime, 
the leader velocity distribution $P(V,t)$
becomes indepedent of $N$ as $N\to \infty$, and satisfies 
to Eq.~(\ref{scaling}) with
\begin{equation}
a_{III}(t)=V_c\,,\quad b_{III}(t)=\left[\frac{(\beta+1)(\beta+2)}
{n_0t \,{\cal C}}\right]^{1/(\beta+2)}\,.
\label{par3}
\end{equation}
The $\beta$-dependent 
universal scaling function is given by 
\begin{eqnarray}
F_{III}^{\beta}(z)&=&(\beta+1)(\beta+2)|z|^{\beta}\int_{|z|}^{\infty} dU\,
e^{-U^{\beta+2}}\nonumber\\
\label{f3}
\end{eqnarray}
for $z \in (-\infty,\, 0]$, and $F_{III}^{\beta}(z)=0$ for $z> 0$.
Its limiting behavior is given by
\begin{equation}
F_{III}^{\beta}(z)\approx \left\{
\begin{array}{ll}
\displaystyle\frac{(\beta+1)}{|z|}\exp[-|z|^{\beta+2}] &\rm{for}\quad z
\rightarrow -\infty\,,\\
& \\
(\beta+1)\;\Gamma\left(\displaystyle\frac{1}{\beta+2}\right)\;|z|^{\beta} &\rm{for}
\quad z \rightarrow 0^-\,.
\end{array}
\right.
\end{equation}
The asymptotic mean velocity of the leader approaches the maximum allowed
value $V_c$ as a power law,
\begin{eqnarray}
\langle V (t)\rangle  &\approx& V_c-\frac{\beta+1}{\beta+2}\;
\Gamma\left(\frac{\beta+3}{\beta+2}\right)\;b_{III}(t)\nonumber\\
&\approx&V_c-\frac{\beta+1}{\beta+2} 
\left[\frac{(\beta+1)(\beta+2)}{{\cal C}}\right]^{1/(\beta+2)}\nonumber\\
&\times& \Gamma\left(\frac{\beta+3}{\beta+2}\right)\;(n_0t)^{-1/(\beta+2)}\,.\nonumber\\
\label{vel3}
\end{eqnarray}

\subsection{Collision statistics of the leader}

We have also computed analytically the mean and the variance 
of $n_c(t,N)$, the number of collisions
undergone by the leader till a given time $t$ and for a given $N\gg 1$. 
As before, one is led to consider the two regimes, 
namely the stationary regime ($t>t^*(N)$) and 
the growing regime ($1 \ll t \ll t^*(N)$).  

In the {\em stationary regime}
the mean and the variance become time-independent,
and they increase logarithmically with $N$ for large $N$. 
The mean behaves as
\begin{equation}
\langle n_c (N)\rangle \approx \xi_i \ln(N)\,,\quad i=I, II, III\,,
\label{ncmst}
\end{equation}
where the universal prefactor $\xi_i$ depends 
on the three classes of $\phi(V)$,
\begin{equation}
\begin{array}{l}
\xi_I = \displaystyle\frac{1}{2}\,,\\
 \\
\xi_{II} = \displaystyle\frac{\beta-2}{2\beta-3}\,,\quad
\mbox{and} \\
\\ 
\xi_{III} = \displaystyle\frac{\beta+2}{2\beta+3}\,.
\end{array}
\label{xist}
\end{equation}
Similarly, the variance in the stationary regime, for large $N$, behaves as
\begin{equation}
{\langle n_c^2 (N)\rangle-\langle n_c (N)\rangle^2} \approx 
{\sigma_i \ln(N)}\,,\quad i=I,II,III\,,
\label{ncvst}
\end{equation}
where
\begin{equation}
\begin{array}{ll}
\sigma_I = \displaystyle\frac{1}{4} \,,&\\
& \\
\sigma_{II} = \displaystyle\frac{(\beta-2)(2\beta^2-6\beta+5)}{(2\beta-3)^3}
\,,&\quad \mbox{and}\\
& \\
\sigma_{III} = \displaystyle \frac{(\beta+2)(2\beta^2+6\beta+5)}{(2\beta+3)^3}
\,.&
\end{array}
\label{sigmast}
\end{equation}
The above results
were first derived analytically in Ref.~\cite{satya1}, using
a different method. This paper provides thus an
alternative derivation.

In the {\em growing regime}, we show that the mean number of collisions increases 
logarithmically with $t$, 
\begin{equation}
\langle n_c (t)\rangle \approx \gamma_i \ln(n_0 t)\,, \quad i=I,II,III\,.
\label{ncas}
\end{equation} 
The universal prefactor $\gamma_i$ is characteristic to 
each of the classes of the
parent-distribution $\phi(V)$, namely
\begin{equation}
\begin{array}{ll}
\gamma_I = \displaystyle\frac{1}{2}\,, &\\
& \\
\gamma_{II} = \displaystyle\frac{\beta-1}{2\beta-3}\,, &\quad \mbox{and}\\
& \\
\gamma_{III} = \displaystyle\frac{\beta+1}{2\beta+3}\,. &
\end{array} 
\label{gamma}
\end{equation}
Moreover, as discussed in Sec.~\ref{collision},
one can also infer the variance of the number of collisions,
\begin{equation}
\langle n_c^2 (t)\rangle-\langle n_c (t)\rangle^2 \approx {\eta_i
\ln(n_0 t)}\,,\quad i=I,II,III\,,
\label{ncas2}
\end{equation}
where
\begin{equation}
\begin{array}{ll}
\eta_I = \displaystyle\frac{1}{4}\,, &\\
& \\
\eta_{II} = \displaystyle\frac{(\beta-1)(2\beta^2-6\beta+5)}{(2\beta-3)^3}\,, &\quad \mbox{and}\\
& \\
\eta_{III} = \displaystyle \frac{(\beta+1)(2\beta^2+6\beta+5)}{(2\beta+3)^3}\,. &
\end{array} 
\label{eta}
\end{equation}
The fact that ${\langle n_c^2 (t)\rangle-\langle n_c (t)\rangle^2}\neq
\langle n_c(t)\rangle$ indicates, contrary to previous claims~\cite{bala1}, that
the collision process in the thermodynamic limit of the 
Jepsen gas is {\em not Poissonian}.

\section{Velocity distribution function of the leader}
\label{pdfsection}

\subsection{General relations}
\label{general}

In order to establish the characteristics of leader's stochastic 
dynamics, we shall follow the type of reasoning  and the convenient 
notations of Ref.~\cite{jarek2}.
Recall that the particles simply exchange velocities upon collisions and 
cannot move across each other; therefore at any time $t$ the leader rides 
the instantaneous
{\em rightmost trajectory} among the free trajectories 
$\{X_i+V_i t,\, i=0,1,...,N\}$. 
As the number of particles on the left-hand side of the leader is a 
conserved quantity, we can identify this instantaneous trajectory 
$(X_p+V_p t)$ of the leader  by imposing:
\begin{equation}
\sum_{i=0, \,i\neq p}^N \theta (X_p+V_p t - X_i-V_i t) = N\,,
\end{equation}
$\theta$ being the Heaviside step function.

These elements are sufficient to find the 
conditional probability distribution of the leader $P_L(X,V,t|V_0)$
at time $t>0$ as
\begin{eqnarray}
&&P_L(X,V,t|V_0)= \langle \sum_{p=0}^N \delta (X-X_p-V_p t)\,\delta(V-V_p)
\nonumber\\
&\times& \delta_{Kr}\left(N,\,\sum_{i=0,\, i\neq p}^N\theta (X_p+V_p t - X_i-V_i
t)\right)\,\rangle\,,\nonumber\\
\label{pl1}
\end{eqnarray}
where  the brackets $\langle ... \rangle$ denote averaging 
over the  initial positions and velocities of 
the gas particles in $[-L,0]$. The initial condition is obviously
\begin{equation}
P_L(X,V,t=0|V_0)=\delta (X) \delta (V-V_0)\,.
\end{equation}

As shown in Appendix A,
one finds
\begin{eqnarray}
&P_L&(X,V,t|V_0)=\oint_{\Gamma}  \frac{dz}{2 \pi i z}
\left[A(z,\,X/t\;|\;L/t)\right]^N\nonumber\\
&&\times\left\{\frac{}{}\delta(X-Vt)\,\delta(V-V_0)\right.\nonumber\\
&&\nonumber\\
&&+\left.
n_0 \phi(V) \theta(Vt-X)\theta(X-Vt+L)\right.\nonumber\\
&&\nonumber\\
&&\times\left.\frac{1+(z-1)\theta(V_0 t -X)}{A(z,\,X/t\;|\; L/t)}\;
\right\}\,.\nonumber\\
\label{main}
\end{eqnarray}
Here $\Gamma$ is the unit-circle in the complex $z$-plane, and
\begin{eqnarray}
&&A(z,\,X/t\;|\; L/t) = 1 + (z-1) \left[\int_{(L+X)/t}^{\infty}
dU\,\phi(U)\right.\nonumber\\
&&\left.+\frac{t}{L}\int_{X/t}^{(L+X)/t} dU\,\left(U - \frac{X}{t}\right) \,\phi(U)\right]\,,
\label{funcA}
\end{eqnarray}
with $n_0=N/L$  the initial density.
One can also check the normalization of  $P_L(X,V,t|V_0)$, see Appendix A.

We now focus on the growing regime, i.e., we fix the time $t$ and take the 
{\em thermodynamic limit} $L\rightarrow \infty$, $N\rightarrow \infty$ 
at constant $n_0$. 
One finds then for the conditional probability distribution of the leader:
\begin{eqnarray}
&&P(X,V,t|V_0)\nonumber\\
&&=\oint_{\Gamma}  \frac{dz}{2 \pi i z} 
\exp\left[n_0t(z-1)\int_{X/t}^{\infty} \,dU\,\left(U - \frac{X}{t}\right)\,\phi(U)\right]\nonumber\\
&&\nonumber\\
&&\times\left\{\delta(X-V_0 t)\,\delta(V-V_0)+
n_0 \phi(V)\theta(Vt-X)\right.\nonumber\\
&&\hspace{3cm}\times\left.
\left[1+(z-1)\theta(V_0t-X)\right]\right\}\,.
\end{eqnarray}
Introducing the function
\begin{equation}
\alpha(W)\equiv\int_W^{\infty}\,dU\,(U-W)\,\phi(U)\,,
\label{alpha}
\end{equation}
one obtains finally:
\begin{eqnarray}
&&P(X,V,t|V_0)=e^{\displaystyle{-n_0t \alpha(V_0)}}\,\delta(X-V_0
t)\,\delta(V-V_0)\nonumber\\
&&+n_0 \phi(V)\,e^{\displaystyle{-n_0t \alpha(X/t)}}\,\theta(Vt-X)\theta(X-V_0t)\,.
\label{complete}
\end{eqnarray}

This result has a simple physical interpretation, based on a ``flux argument". 
Let $V$ be the instantaneous 
velocity of the 
leader at time $t$. The leader's trajectory can get bypassed only by 
the trajectories  with higher initial velocities 
$U>V$; then the rate at which the trajectory of the leader gets bypassed 
is proportional to 
the flux of particles trajectories of slopes higher than $V$. 
This flux is clearly proportional
to the particle density $n_0$ and to the relative velocity 
$(U-V)\theta(U-V)$. Therefore, 
$n_0\alpha(V)=n_0 \displaystyle \int_{V}^{\infty} dU (U-V)\phi (U)$
represents the total instantaneous rate at which the leader's trajectory 
with velocity $V$ gets bypassed by other trajectories. 

The first term in the r.h.s. of Eq.~(\ref{complete}) represents 
the probability that the leader's trajectory has never been bypassed till 
time $t$. One notices that this probability is exponentially-decreasing with time. 
In this case, the final velocity is $V_0$ 
and the final position is $V_0t$, thus explaining the two $\delta$ factors. 
Indeed, as seen above,
in an infinitesimal time $dt$ the trajectory of the leader with velocity $V_0$ 
gets hit with probability
$n_0 \alpha (V_0) dt$, and so it does not get hit with probability 
$1-n_0 \alpha (V_0)dt$. 
As such, the probability for the leader to keep its initial trajectory till time $t$ is 
$[1-n_0\alpha (V_0)dt]^{\displaystyle t/dt} \rightarrow \exp[-n_0t\alpha (V_0)]$.

\begin{figure}
\begin{center}
\includegraphics[width=0.88\columnwidth]{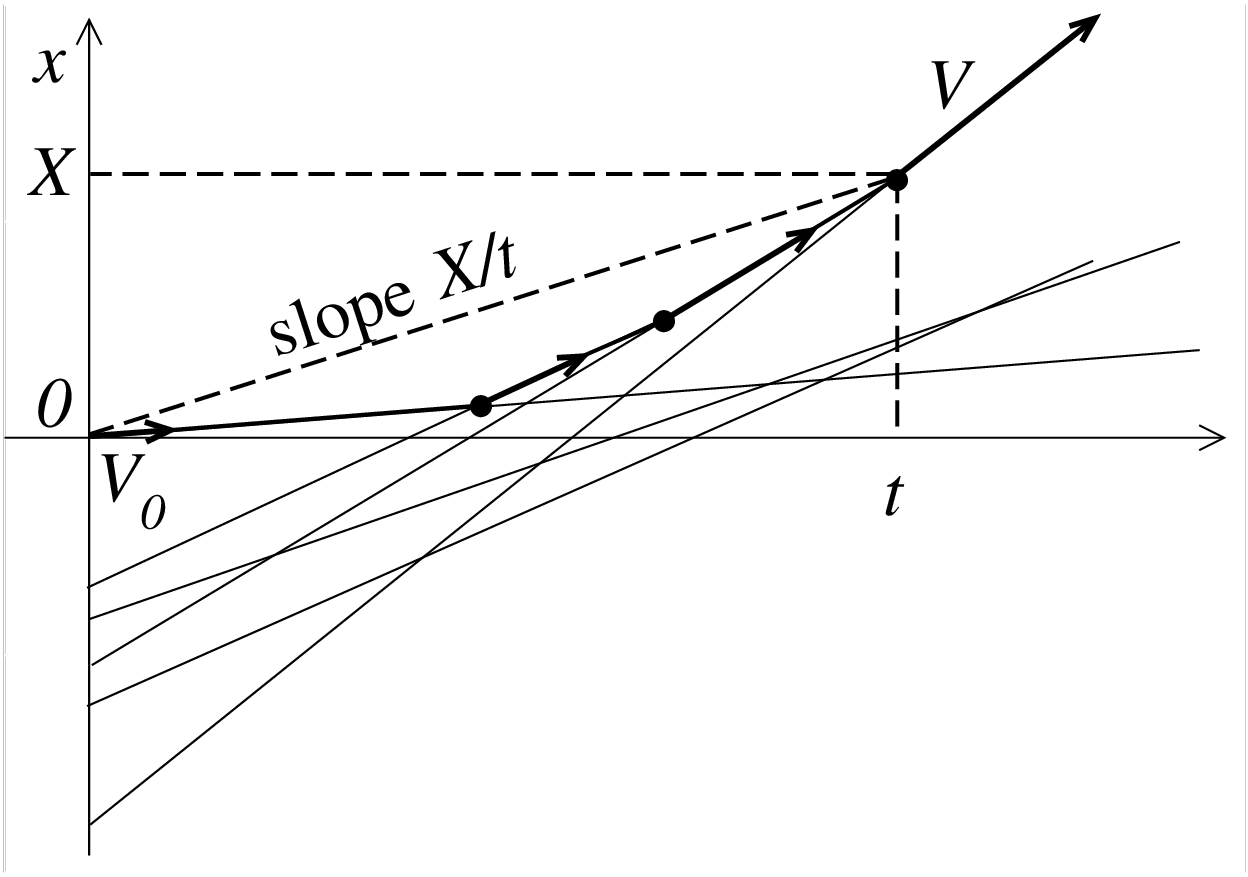}
\caption{\label{figure3} Illustration of the ``flux argument" in deducing the
conditional probability distribution function $P(X,V,t|V_0)$ of the leader in the
thermodynamic limit (see the main text).}
\end{center}
\end{figure}

Let us turn now to the second term of the r.h.s. of Eq.~(\ref{complete}), 
which takes into account all the situations when the velocity of the leader 
got modified through collisions. 
The leader is at $X$ at time $t$, with velocity $V$. For this event 
to happen, no trajectory must hit the straight line of slope $X/t$ 
till time $t$, see Fig.~\ref{figure3}. 
All possible trajectories of the leader must lie below 
this line, and (according to the argument above) this happens 
with probability $\exp[-n_0t\alpha(X/t)]$. Now, out of all the
trajectories satisfying this criterion, we are interested only 
in those that actually hit the line of slope $X/t$ exactly at time $t$, 
and exactly with velocity $V$; this fraction is $n_0\phi(V)$. Thus, 
the total probability for the leader to be at $X$ with velocity $V$
is $n_0\phi(V) \exp[-n_0t\alpha(X/t)]$ 
(the two $\theta$ functions are obvious).

The conditional coordinate distribution for the leader 
is obtained by integrating 
$P(X,V,t|V_0)$ over $V$:
\begin{eqnarray}
P(X,t|V_0)&=&e^{\displaystyle{-n_0t \alpha(V_0)}}\,\delta(X-V_0t)\nonumber\\
&+&n_0e^{\displaystyle{-n_0t \alpha(X/t)}}
\int_{X/t}^{\infty} \,dV \phi(V) \,\theta(X-V_0t)\nonumber\\
&&\nonumber\\
&=&\frac{\partial}{\partial X}\left\{
e^{\displaystyle{-n_0t \alpha(X/t)}}\,\theta(X-V_0t)\right\}\,,
\label{position}
\end{eqnarray}
and a detailed discussion of its long-time 
properties, corresponding to the diffusive regime for the particle,  
can be found in Ref.~\cite{lebowitz1}.

We now consider exclusively the 
stochastic behavior of the 
velocity of the leader.  The {\em conditional} velocity distribution
is obtained by integrating $P(X,V,t|V_0)$ over $X$:
\begin{eqnarray}
P(V,t|V_0)&=&e^{\displaystyle{-n_0t \alpha(V_0)}}\,\delta(V-V_0)\nonumber\\
&+& n_0 t\phi(V)\,
\int_{V_0}^{V}\,dW\,e^{\displaystyle{-n_0t \alpha(W)}}\,\theta(V-V_0)\,.\nonumber\\
\label{velocity}
\end{eqnarray}
Moreover, by averaging over the initial velocity $V_0$ of the leader, one obtains the 
{\em velocity distribution}:
\begin{eqnarray}
P(V,t)&=&\int_{0}^{\infty} dV_0\, \phi(V_0)P(V,t|V_0)\nonumber\\
&=&\phi(V) e^{\displaystyle{-n_0t\alpha(V)}}\nonumber\\
&+&n_0 t\phi(V)\int_{0}^{V}dV_0
\phi(V_0)\int_{V_0}^V dW\,e^{\displaystyle{-n_0t\alpha(W)}}\,.\nonumber\\
\label{pdf}
\end{eqnarray}  
Based on the properties of the function $\alpha(W)$ (see Appendix B),
one reaches finally an expression that proves to be more convenient 
for further analysis,
\begin{equation}
P(V,t)=\phi(V)e^{\displaystyle{-n_0t\mu}}+n_0t \phi(V)\int_0^V dW\, 
e^{\displaystyle{-n_0t\alpha(W)}}\,.
\label{pdf0}
\end{equation}
Here $\mu>0$ is the first moment of $\phi(V)$,
\begin{equation}
\mu=\int_0^{\infty} dV\;V \phi(V)\,.
\end{equation}

\subsection{Long-time behavior of $P(V,t)$}

Let us inspect Eq.~(\ref{pdf0}) in the limit $t\gg (n_0\mu)^{-1}$. 
The first term in the r.h.s.
is exponentially decreasing in time, so it can be neglected 
in this long-time limit.
In the second term,
due to the presence of the exponential $\exp[-n_0t\alpha(W)]$, 
the main contribution to the integral will come from the $W$ sector for which 
$\alpha(W)$ is small; in view of the property $(i)$ in Appendix B, this happens for
large values of $W$, so one can write with a good approximation:
\begin{equation}
P(V,t) \approx n_0t \phi(V) \int_0^{\infty} dW \exp[-n_0t\alpha^{\rm{as}}(W)]\,.
\label{pdflongtime}
\end{equation}
Here ``as" designates the asymptotic, large-$W$ behavior of $\alpha(W)$,
which, in view of the definition~(\ref{alpha}), is determined by 
the asymptotic behavior of  $\phi(V)$. We are thus led to consider 
the tail of the distribution $\phi(V)$,
according to the three classes discussed in Sec.~\ref{summary}.\\
{\em Class I, Eq.~(\ref{class1})}: In this case one finds
\begin{equation}
\alpha^{\rm as}(W) \approx \frac{{\cal A}}{\delta^2}\;W^{\beta+2(1-\delta)}\;\exp(-W^{\delta})\,.
\label{alpha1}
\end{equation}
As shown in Appendix C, this leads to the scaling form~(\ref{scaling}) with the
scaling function $F_I$ in Eq.~(\ref{f1}) and the scaling parameters~(\ref{par1}).\\
{\em Class II, Eq.~(\ref{class2})}: One has
\begin{equation}
\alpha^{\rm as}(W) \approx \frac{{\cal B}}{(\beta-1)(\beta-2)}\;\frac{1}{W^{\beta-2}}\,.
\label{alpha2}
\end{equation}
Introducing it in Eq.~(\ref{pdflongtime}) and using a simple change of variable
\begin{equation}
\left[\frac{n_0t{\cal B}}{(\beta-1)(\beta-2)}\right]^{-1/(\beta-2)}\;V=z\,,
\end{equation}
one obtains the scaling form~(\ref{scaling}) with 
$F_{II}^{\beta}$  as in Eq.~(\ref{f2}) and the scaling parameters~(\ref{par2}).\\
{\em Class III, Eq.~(\ref{class3})}: Finally, for distributions with 
finite support $\phi(V)$
one has
\begin{equation}
\alpha^{\rm as}(W) \approx \frac{{\cal C}}{(\beta+1)(\beta+2)}{(V_c-W)^{\beta+2}}\,,
\quad W \leqslant V_c\,.
\label{alpha3}
\end{equation}
The change of variable
\begin{equation}
\left[\frac{n_0t{\cal C}}{(\beta+1)(\beta+2)}\right]^{1/(\beta+2)}\,(V_c-W)=-z\,,
\end{equation}
leads to the scaling form~(\ref{scaling}) with 
$F_{III}^{\beta}$  given by Eq.~(\ref{f3}) 
and the scaling parameters in~(\ref{par3}).\

\subsection{Mean velocity of the leader}

Using Eq.~(\ref{pdf}) and the properties (B3)-(B6) of $\alpha(V)$, 
followed by a double
integration by parts, the mean velocity of the leader can be written  as
\begin{eqnarray}
&&\langle V(t) \rangle=\int_{0}^{\infty}
dV\;V\,P(V,t)=\mu\,e^{-n_0t\mu}\nonumber\\
&&+n_0t\int_0^{\infty}
dV\;\left[\alpha(V)-V\frac{d\alpha(V)}{dV}\right]\,e^{-n_0t\alpha(V)}\,.\nonumber\\
\label{avvel}
\end{eqnarray}
Or, with a change of variable
\begin{equation}
Z=n_0t\alpha(V)
\label{changevar}
\end{equation}
the above integral becomes
\begin{equation}
\langle V (t)\rangle =\mu\,e^{-n_0t\mu}+\int_0^{n_0t\mu}
dZ\;\left(\frac{\alpha(V)}{|d\alpha(V)/dV|}+V\right)\,e^{-Z}\,.
\label{meanval}
\end{equation}
The quantity $\left(\displaystyle\frac{\alpha(V)}{|d\alpha(V)/dV|}+V\right)$ 
has to be expressed as a
function of $Z$ using Eq.~(\ref{changevar}).
 
In the {\em long-time limit} of $t \gg (n_0\mu)^{-1}$, the first term in the 
r.h.s. of Eq.~(\ref{meanval}) becomes negligible, and
the dominant contribution to the integral
in the second term comes only from the large-$V$ sector. 
Therefore,
\begin{equation}
\langle V (t)\rangle \approx n_0t \int_0^{\infty} dZ
\left(\frac{\alpha^{{\rm as}}(V)}{|d\alpha^{{\rm as}}(V)/dV|}+V\right)\,e^{-Z}\,,
\end{equation} 
with $Z=n_0t \alpha^{{\rm as}}(V)$, 
so the long-time behavior of the mean velocity is determined by the 
tail of the parent distribution $\phi(V)$. Therefore, 
its expression is the same for 
all the distribution $\phi(V)$ belonging to one of the 
three classes described
above, and is given, respectively, by the Eqs.~(\ref{vel1}), (\ref{vel2}), and
(\ref{vel3}) above.

Note that these long-time expressions of the mean leader velocity 
can also be obtained
directly by using the specific scaling 
forms~(\ref{scaling}) of $P(V,t)$.

\section{Collision statistics}
\label{collision}

\subsection{Growing regime}

We are interested in the statistics of the number of collisions $n_c$
that the leader undergoes till time $t$ in the growing regime. 
As before, we then fix the time $t$ and take first the thermodynamic limit
$N\to \infty$, $L\to \infty$ keeping the density $n_0$ fixed.
 If $V$ is leader's velocity 
at time $t$, then the probability that it undergoes one collision in the 
small interval $dt$ is $n_0\alpha(V)dt$, see Sec.~\ref{general}. 
Taking the average over the distribution
$P(V,t)$ of $V$, one obtains:
\begin{equation}
\frac{d\langle n_c(t)\rangle}{dt}=
n_0 \int_0^{\infty}dV\;\alpha(V) P(V,t)= n_0\langle \alpha(V) \rangle\,.
\label{nc}
\end{equation}
The long-time behavior is obtained by considering in the above equation the 
asymptotic form~(\ref{scaling}) of $P(V,t)$. One notices that the most
important contribution to the integral over $V$ comes from the region 
$V\approx a_i(t)\,,i=I,II,III$.
For the three classes of parent
distributions $\phi(V)$ one obtains, respectively:\\
{\em Class I}:
\begin{eqnarray}
&&\frac{d\langle n_c(t)\rangle}{dt}\nonumber\\
&&\approx n_0 \int_0^{\infty}dV \,\frac{{\cal A}}{\delta^2}\,
V^{\beta+2-2\delta}e^{-V^{\delta}}\,\frac{1}{b_I(t)}\,F_I\left(\frac{V-a_I(t)}{b_I(t)}\right)\nonumber\\
&&\approx\frac{1}{t \;\delta^2}\int_0^{\infty}dV\;
\exp\left[-V^{\delta}+(\beta+2-2\delta)\ln V+\right.\nonumber\\
&&\left.+\ln (n_0t{\cal
A})\right]\,\frac{1}{b_I(t)}
\,F_I\left(\frac{V-a_I(t)}{b_I(t)}\right)\nonumber\\
&&\approx\frac{1}{t}\int_{-\infty}^{\infty} dz\;e^{-z}\,F_I(z) \approx
\frac{1}{2\;t}\,.
\end{eqnarray}
{\em Class II}:
\begin{eqnarray}
\frac{d\langle n_c(t)\rangle}{dt} &\approx& 
\frac{1}{t}\int_0^{\infty}dz\;z^{2-\beta}
\,F_{II}^{\beta}(z)\approx \left(\frac{\beta-1}{2\beta
-3}\right)\,\frac{1}{t}\,.\nonumber\\
\end{eqnarray}
{\em Class III}:
\begin{eqnarray}
\frac{d\langle n_c(t)\rangle}{dt} &\approx&
\frac{1}{t}\int_{-\infty}^{0}dz\;(-z)^{\beta+2}\,F_{III}^{\beta}(z)
\approx \left(\frac{\beta+1}{2\beta +3}\right)\,\frac{1}{t}\,.\nonumber\\
\end{eqnarray}
These results lead obviously to the $\propto \ln(n_0 t)$ asymptotic growth
of $\langle n_c(t)\rangle$, as described by Eqs.~(\ref{ncas})-(\ref{gamma}).

The mean square number of collisions till time $t$ 
is given by the following integral expression
\begin{eqnarray}
\langle n_c^2(t)\rangle &=& 2 \int_0^t
dt_1\int_{t_1}^t dt_2\int_0^{\infty}
dV_1\int_{V_1}^{\infty}dV_2\int_{V_1}^{V_2} dU\nonumber\\
&\times& n_0 (U-V_1)\phi(U) \,P(V_2,t_2|U,t_1) n_0\alpha(V_2)\,.\nonumber\\
\label{square}
\end{eqnarray}
The kernel of this five-fold integral corresponds to:\\ 
(i) having one collision in the interval $dt_1$ around $t_1$,
provided that the leader has velocity $V_1$ at $t_1$; (ii) as a result
of this first collision, an instantaneous change in leader's 
velocity from $V_1$ to $U>V_1$;
(iii) a second collision in the interval $dt_2$ around $t_2$, 
provided that the leader has velocity $V_2\geqslant U$ at $t_2$ 
($t_2 > t_1$)
and that it had the velocity $U$ at $t_1^+$. The corresponding conditional
probability density $P(V_2,t_2|U,t_1)$ can be easily computed using
Eq.~(\ref{complete}) and integrating over the spatial coordinates,
and it is found to be:
\begin{eqnarray}
&&P(V_2,t_2|U,t_1)=e^{\displaystyle{-n_0(t_2-t_1) \alpha(U)}}\,\delta(V_2-U)\nonumber\\
&+& n_0 (t_2-t_1)\phi(V_2)\,
\int_{U}^{V_2}\,dW\,e^{\displaystyle{-n_0(t_2-t_1) \alpha(W)}}\,\theta(V_2-U)\,.\nonumber\\
\label{condvelocity}
\end{eqnarray}
Finally, one has to average over all the realizations of leader's 
trajectory that fulfill conditions (i)-(iii).

From Eqs.~(\ref{square}), (\ref{condvelocity}), and (\ref{ncas})-(\ref{gamma}),  
one can infer the long-time behavior $\propto \ln(n_0 t)$ of the variance of $n_c$,
as resumed in Eqs.~(\ref{ncas2})-(\ref{eta}). This calculation is rather tedious
and we have carried it out explicitly only in a special case when $\phi(V)=\exp(-V)$
(see Sec.~V). However, the results for other cases can be inferred by matching 
the late-time growing regime results with that of the stationary regime results
derived in Ref. \cite{satya1} at the crossover time $t=t^*(N)$. This is explained
in detail in the next subsection.

\subsection{Stationary regime and the crossover time}

As already mentioned in the Introduction, the Jepsen gas for
a finite $N$ is completely equivalent to the model of an evolutionary
dynamics for a quasispecies model introduced in~\cite{krug,jain}.
The collision statistics of the leader in the stationary regime
has been studied before both numerically~\cite{krug,jain} 
and analytically~\cite{satya1}, 
and the results in Eqs.~(\ref{ncmst}) and (\ref{ncvst}) were derived. Here
we will derive these results through a different method, that 
will also allow to estimate the crossover time $t^*(N)$ for the three classes of $\phi(V)$.

Recall that the crossover time $t^*(N)$ is 
the time at which the leader aquires its
final, maximum  velocity. So, at $t=t^*(N)$ the typical time-dependent
velocity of the leader in the growing regime matches with 
the typical value of $V_{\rm max}=\rm{max}(V_0, V_1, ..., V_N)$.
To estimate $V_{\rm max, typ}$, we recall that
the $V_i$-s are i.i.d random variables each drawn from
$\phi(V)$, and therefore, for $N\gg 1$:  
\begin{eqnarray}
&&\rm{Prob}(V_{{\rm{max}}}\leqslant {\cal V})
=\prod_{i=0}^N \rm{Prob}(V_i \leqslant {\cal V})\nonumber\\
&&=\left[\int_0^{{\cal
V}}dU\,\phi(U)\right]^{N+1}
=\left[1-\int_{{\cal V}}^{\infty}dU\,\phi(U)\right]^{N+1}\nonumber\\
&&\approx \exp\left[-N\int_{{\cal V}}^{\infty}dU\,\phi(U)\right] = 
\exp \left[N\alpha'({\cal V})\right]\,,\nonumber\\
\label{maxval}
\end{eqnarray}
where the function $\alpha(z)$ is defined in Eq.~(\ref{alpha}).

{\em Class I}: In this case, Eq.~(\ref{maxval}) generates the 
cumulative of the Fisher-Tippett-Gumbel p.d.f.,
\begin{eqnarray}
&&\rm{Prob}(V_{\rm{max}}\leqslant {\cal V})\nonumber\\
&&\approx
\exp\left[-\exp\left(-\frac{{\cal V}-{\tilde a}_I(N)}{{\tilde b}_I(N)}\right)\right]\,,
\label{gumbel}
\end{eqnarray}
where ${\tilde a}_I(N)$ and ${\tilde b}_I(N)$ are given in Eq.~(\ref{sc1}).
So in this case the typical value of $V_{\rm max}$ is
$V_{\rm {max, typ}} \approx {\tilde a}_I(N)$. On the other hand, as indicated by
Eqs.~(\ref{scaling}), (\ref{par1}), (\ref{vel1}), the typical leader velocity
at time $t$ is roughly $V_{{\rm typ}}(t)\approx a_I(t)$. 
Matching $V_{\rm {max,typ}}=V_{\rm{typ}}(t^*(N))$ 
gives the crossover time,
\begin{equation}
t^*(N) \approx \frac{\delta}{n_0}N\,.
\label{cross1}
\end{equation}
{\em Class II}: In this case, one obtains the cumulative of the
Fr\'echet p.d.f.,
\begin{equation}
\rm{Prob}(V_{\rm{max}}\leqslant {\cal V})\approx
\exp\left[-\frac{N {\cal B}}{\beta-1}{\cal V}^{1-\beta}\right]\,,
\end{equation}
and hence
\begin{equation}
{V}_{\rm{max,typ}} \approx  N^{1/(\beta-1)}\,.
\end{equation}
On the other hand, according to 
Eqs.~(\ref{scaling}), (\ref{par2}), (\ref{vel2}), the typical leader velocity
\begin{equation}
V_{\rm{typ}}(t) \approx \left[\frac{{\cal B}n_0t}{(\beta-1)(\beta-2)}\right]^{1/(\beta-2)}\,.
\end{equation}
Then the crossover time is estimated as:
\begin{equation}
t^*(N) \approx N^{(\beta-2)/(\beta-1)}\,.
\label{cross2}
\end{equation}
{\em Class III}: Finally, one has in this case the cumulative of 
the Weibull p.d.f.,
\begin{eqnarray}
&&\rm{Prob}(V_{\rm{max}}\leqslant {\cal V}]
\approx \exp\left[-\frac{N {\cal C}}{\beta+1}
(V_c-{\cal
V})^{\beta+1}\right]\,,
\end{eqnarray}
indicating
\begin{equation}
V_c-V_{\rm {max,typ}} \approx  N^{-1/(\beta+1)}\,.
\end{equation}
The typical leader velocity is given, 
through Eqs.~(\ref{scaling}), (\ref{par3}), and (\ref{vel3}), by
\begin{equation}
V_c-V_{\rm{typ}}(t) \approx \left[\frac{{\cal C}n_0t}
{(\beta+1)(\beta+2)}\right]^{-1/(\beta+2)}\,,
\end{equation}  
and thus the crossover time is
\begin{equation}
t^*(N) \approx N^{(\beta+2)/(\beta+1)}\,.
\label{cross3}
\end{equation}

Substituting Eqs.~(\ref{cross1}), (\ref{cross2}), (\ref{cross3}) in
Eq.~(\ref{ncas}), one can then compute  
the saturation value of the mean collision number 
$\langle n_c^*(N)\rangle=\langle n_c(t=t^*(N))\rangle$ as a 
function of $N$ as stated in Eq. (\ref{ncmst}). We thus recover, through this
alternative method,
the results derived in Ref.~\cite{satya1}.

Concerning the variance, the results in Eqs.~(\ref{ncvst}), (\ref{sigmast})
for the stationary regime were derived exactly in Ref.~\cite{satya1}. 
In order that the variance in the growing regime matches with that 
in the stationary regime at the crossover time $t=t^*(N)$, 
it follows immediately that in the growing regime the variance at 
late times must behave as in Eqs.~(\ref{ncas2}), (\ref{eta}). 
We have verified 
this prediction by direct calculation
for the special case $\phi(V)=\exp(-V)$
(see the next Section). 
However, although desirable, a direct calculation 
of the variance in the growing regime for the 
other cases,
using the method outlined in the previous subsection, 
seems too tedious.

\section{Exact results for all time in a special case}
\label{example}

A particular example that can be studied in full analytical 
detail for all $t$ is
\begin{equation}
\phi(V)=e^{-V} \quad (V\geqslant 0)\,,
\end{equation}
which pertains to Class I with $\delta=1$.
By straightforward calculations one obtains from Eqs.~(\ref{pdf0}),
(\ref{avvel}),
and (\ref{nc}), respectively:\\
(i) The probability distribution function for the velocity of the leader
\begin{equation}
P(V,t)=e^{-(n_0t+V)}+n_0te^{-V}\left[\mbox{Ei}(-n_0t)-\mbox{Ei}(-n_0te^{-V})\right]\,,
\label{pvexp}
\end{equation}
with the scaling form~(\ref{scaling}) corresponding to the function~(\ref{f1})
with the parameters
\begin{equation}
a_I(t)=\ln(n_0t)\,,\quad b_I=1\,.
\end{equation}
(ii) The mean velocity is therefore
\begin{equation}
\langle V(t)\rangle =\ln (n_0t)+1 +C -\mbox{Ei}(-n_0t)\,,
\end{equation}
which for long times is dominated by the logarithmic term.\\
(iii) The mean value of the number of collisions the leader undergoes till
time $t$:
\begin{equation}
\langle n_c(t)\rangle =\frac{1}{2}\left[\ln (n_0t) +C -\mbox{Ei}(-n_0t)\right]\,,
\label{ncexp}
\end{equation}
with the logarithmic asymptotic increase.\\
(iv) Finally, the calculation of the variance of the number 
of collisions is rather lengthy (see Appendix~\ref{msdcol}), but its long-time
behavior is simply given by
\begin{equation}
\langle n_c^2(t)\rangle - \langle n_c(t)\rangle^2 \approx \frac{1}{4}\,\ln(n_0 t)\,,
\label{nc2exp}
\end{equation}
in agreement with Eqs.~(\ref{ncas2}) and (\ref{eta}).
 
\section{Conclusions}
\label{end}

In this paper we have studied the extremal dynamics in a 
one-dimensional Jepsen gas of $(N+1)$ particles, initially confined
in a box $[-L,0]$ with uniform density and with each particle 
having an independently distributed
initial positive velocity drawn from an arbitrary distribution $\phi(V)$. 
We have computed analytically the velocity 
distribution
of the leader (or the rightmost particle) at time $t$,
and also the mean and the variance
of the number of collisions undergone by the leader up to time $t$. 
We have shown that
for a given $N \gg 1$, there is a crossover time $t^*(N)$ that separates a
stationary regime ($t>t^*(N)$) from a growing regime ($1 \ll t \ll t^*(N)$).
While in the stationary regime, the leader velocity becomes time-independent
and follows the standard extremal laws of i.i.d random variables, it has
novel universal scaling behavior in the growing regime. The associated
scaling functions in the growing regime belong to 
three different universality classes depending
only on the tail of $\phi(V)$, and they have been computed 
explicitly in Eqs.~(\ref{f1}), (\ref{f2}), and (\ref{f3}).
These dynamical extremal scaling functions are manifestly different
from the standard EVS scaling functions of i.i.d random variables. 

Similarly, we have shown that in the growing regime, the mean and 
the variance of the number of collisions 
of the leader
up to time $t$ increases logarithmically with $t$, with universal prefactors
that were computed explicitly in Eqs.~(\ref{gamma}) and (\ref{eta}).
Also, as a by-product, we have provided an alternate derivation of the 
stationary regime results for the collision statistics 
(mean and the variance) of the leader that
were obtained in~\cite{satya1} through a completely different approach.
While in this paper we were able to compute only the mean and the variance
of the number of collisions, it would be interesting to compute the full
distribution of the number of collisions up to time $t$, 
and to compare it with the previously incorrectly suggested
Poisson distribution~\cite{bala1}, which remains a
challenging open problem.

We have computed here the velocity distribution and the statistics 
of the number
of collisions {\em separately} in the growing regime and in the 
stationary regime. It would be
interesting to compute the exact crossover functions 
that interpolate between the two regimes.
For example, for the velocity distribution of the leader 
this crossover scaling function can, in principle, 
be computed from our general result in 
Eq.~(\ref{main}), which is valid for all times $t$ and all
values of  $N$.

The model studied here can also be considered as a simple 
toy model of biological evolution~\cite{krug,jain,satya1}, on
which some results, numerical and analytical, were known before
but they were mostly restricted in the stationary regime~\cite{krug,jain,satya1}. 
For example,
the number of overtaking events of the leader, i.e., 
the number of punctuation
events till the emergence of the best fitted species 
were studied before~\cite{krug,jain,satya1}.
However, the authors in Ref.~\cite{jain} also studied analytically
the distribution of the `label' of the fittest 
species in the growing regime, and by matching 
the typical 
leader's label at time $t$
with that of the final leader, 
they were able to extract the crossover time 
$t^*(N)$. In this paper, we have studied a complimentary quantity in the growing
regime, namely the distribution of the `fitness' (velocity) of the fittest species.
The crossover time $t^*(N)$ extracted from both of these distributions are in 
agreement with each other.   
The method presented in this paper 
may also be useful to study the dynamics 
of other interesting observables in the context of 
biological evolution, such as, for example, the 
persistence of the leader genotype, and
the distribution of the time interval between 
two successive punctuation events.

\begin{acknowledgments}
I.B. is grateful to Profs. 
Christian Van den Broeck, Michel Droz, and Zolt\'an R\'acz for
very useful discussions during the early stages of this work  
and acknowledges partial support from the Swiss National 
Science Foundation. S.N.M. thanks the hospitality of the Isaac Newton
Institute (Cambridge, UK) where part of this work was done.

\end{acknowledgments}

\appendix

\section{Derivation of Eq.~(\ref{main}) and the normalization of $P_L(X,V,t|V_0)$}

We present below the main steps of the derivation of Eq.~(\ref{main}),
see also~\cite{jarek2}. 
In order to pursue the calculations, it is convenient to use the 
following integral representation of the Kronecker delta:
\begin{equation}
\delta_{Kr}(p,q)=\oint_{\Gamma} \frac{dz}{2 \pi i z} \,z^{p-q}\,,
\label{intkr}
\end{equation}  
where $p,q$ are integers, and $\Gamma$ is the unit 
circle centered at the origin in the complex $z$-plane.
Using this representation in
Eq.~(\ref{pl1}), it results:
\begin{eqnarray}
&&P_L(X,V,t|V_0)\nonumber\\
&=&\langle \sum_{p=0}^N \delta(X-X_p-V_pt)\delta(V-V_p)\oint_{\Gamma}\frac{dz}{2 \pi
i z}\,z^N \nonumber\\
&&\hspace{3cm}\times \,\prod_{j=0,j\neq p}^N\,z^{-\theta(X_p+V_pt-X_j-V_jt)}\rangle\nonumber\\
&=&\oint_{\Gamma}\frac{dz}{2\pi i z} z^N \left\{\delta(X-V_0t)\delta(V-V_0)
\langle \prod_{j=1}^{N}z^{-\theta(V_0t-X_j-V_jt)}\rangle \right.\nonumber\\
&+&\left. \langle
\sum_{p=1}^{N}\delta(X-X_p-V_pt)\delta(V-V_p)z^{-\theta(X_p+V_pt-V_0t)}\right.\nonumber\\
&&\left.\hspace{3cm}\times\prod_{j=1,j\neq p}^Nz^{-\theta(X_p+V_pt-X_j-V_jt)}\rangle \right\}\,.\nonumber\\
&&
\label{app1}
\end{eqnarray}
Here $\theta(...)$ is the Heaviside unit step function.
Due to the initial statistical independence of the particles, one has:
\begin{eqnarray}
\langle \prod_{j=1}^{N}z^{-\theta(V_0t-X_j-V_jt)}\rangle&=&
\langle z^{-\theta(V_0t-X_j-V_jt)}\rangle^N\nonumber\\
&=&\langle z^{-\theta(X-X_j-V_jt)}\rangle^N\,,
\label{app2}
\end{eqnarray}
where the last equality results because of
the factor $\delta(X-V_0t)$ in the 
corresponding term of the equation~(\ref{app1}).
Also:
\begin{eqnarray}
&&\langle \sum_{p=1}^N
\delta(X-X_p-V_pt)\delta(V-V_p)z^{-\theta(X_p+V_pt-X_0-V_0t)}\nonumber\\
&&\hspace{3cm}\times
\prod_{j=1, j\neq p}^N z^{-\theta(X_p+V_pt-X_j-V_jt)}\rangle \nonumber\\
&&\nonumber\\
&&=N\langle \delta(X-X_p-V_pt)\delta(V-V_p)\rangle z^{-\theta(X-V_0t)}\nonumber\\
&&\hspace{3cm}\times \langle z^{-\theta(X-X_j-V_jt)}\rangle^{N-1}\,.
\label{app3}
\end{eqnarray}

One has to compute then
\begin{eqnarray}
&&\langle z^{-\theta(X-X_j-V_jt)}\rangle\nonumber\\
&&=\frac{1}{L}\int_{-L}^0 dX_j\int_{0}^{\infty}dV_j\phi(V_j)z^{-\theta(X-X_j-V_jt)}\nonumber\\
&&\nonumber\\
&&=\frac{1}{L}\int_{-L}^0dX_j\left[z^{-1}\int_{0}^{\frac{X-X_j}{t}}dV_j\phi(V_j)+
\int_{\frac{X-X_j}{t}}^{\infty}dV_j\phi(V_j)\right]\nonumber\\
&&\nonumber\\
&&=z^{-1}+(1-z^{-1})\frac{1}{L}\int_{-L}^0 dX_j\int_{\frac{X-X_j}{t}}^{\infty}dV_j\phi(V_j)\nonumber\\
&&=z^{-1}+(1-z^{-1})\left[\int_{\frac{X+L}{t}}^{\infty}dV_j\phi(V_j)\right.\nonumber\\
&&\hspace{3cm}+\left.
\frac{1}{L}\int_{\frac{X}{t}}^{\frac{X+L}{t}}dV_j\phi(V_j)(V_jt-X)\right]\nonumber\\
&&\nonumber\\
&&\equiv z^{-1} A(z,X/t\;|\;L/t)\,,
\label{app4}
\end{eqnarray}
where
$A(z,X/t\;|\;L/t)$ is given by Eq.~(\ref{funcA}).
Note also that
\begin{eqnarray}
&&\langle \delta(X-X_p-V_pt)\delta(V-V_p)\rangle \nonumber\\
&&\nonumber\\
&=&\frac{1}{L}\int_{-L}^0dX_p\int_{-\infty}^{\infty}dV_p\phi(V_p)\delta(X-X_p-V_pt)\delta(V-V_p)\nonumber\\
&&\nonumber\\
&=&\frac{1}{L}\int_{-L}^0dX_p\delta(X-X_p-Vt)\phi(V)\nonumber\\
&&\nonumber\\
&=&
\frac{1}{L}\phi(V)\theta(Vt-X)\theta(X-Vt+L)\,,
\label{app5}
\end{eqnarray}
and also
\begin{eqnarray}
z^{-\theta(X-V_0t)}&=&z^{-1}+(1-z^{-1})\theta(V_0t-X)\nonumber\\
&=&z^{-1}\left[1+(z-1)\theta(V_0t-X)\right]\,.
\label{app6}
\end{eqnarray}
Finally, replacing the results~(\ref{app2})--(\ref{app6}) in Eq.~(\ref{app1}),
one obtains the expression~(\ref{main}) of the conditional probability density for
leader's coordinate and velocity. One can also check the normalization of this one.
Indeed, 
consider
\begin{eqnarray}
{\cal{N}}&=&\int_{-\infty}^{\infty}dX\int_{-\infty}^{\infty}dV
P_L(X,V,t|X_0=0,V_0)\nonumber\\
&=&\oint_{\Gamma}\frac{dz}{2 \pi i z}\left\{\frac{}{}[A(z,V_0\;|\;L/t)]^N
+\frac{N}{L}\int_{-\infty}^{\infty}dX
\int_{\frac{X}{t}}^{\frac{X+L}{t}}dV\right.\nonumber\\
&\times&\left.\phi(V)\,
[A(z,X/t\;|\;L/t)]^{N-1} \frac{}{}[1+(z-1)\theta(V_0t-X)]\right\}
\nonumber\\
&&\nonumber\\
&=&[A(0,V_0\;|\;L/t)]^N
+\frac{N}{L}\int_{-\infty}^{\infty}dX\int_{\frac{X}{t}}^{\frac{X+L}{t}} 
dV\nonumber\\
&\times&\phi(V)[A(0,X/t\;|\;L/t)]^{N-1} \theta(X-V_0t)\,,
\end{eqnarray}
where for the last equality we used the theorem of residues.
But:
\begin{eqnarray}
&&\frac{\partial}{\partial
X}[A(0,V_0\;|\;L/t)]^N\nonumber\\
&&=\frac{N}{L}[A(0,X/t\;|\;L/t)]^{N-1}
\;\int_{\frac{X}{t}}^{\frac{X+L}{t}}dU\phi(U)\,,
\end{eqnarray}
and therefore:
\begin{eqnarray}
{\cal{N}}&=&[A(0,V_0\;|\;L/t)]^N+\int_{-\infty}^{\infty}dX\theta(X-V_0t)\nonumber\\
&&\hspace{2cm}\times\frac{\partial }{\partial X}[A(0,X/t\;|\;L/t)]^{N}\nonumber\\
&&\nonumber\\
&=&[A(0,V_0\;|\;L/t)]^N+\int_{V_0t}^{\infty}dX\frac{\partial }{\partial X}[A(0,X/t\;|\;L/t)]^{N}\nonumber\\
&&\nonumber\\
&=&\lim_{X\rightarrow \infty}[A(0,X/t\;|\;L/t)]^{N}=1\,,
\end{eqnarray}
which proves the normalization of 
$P_L(X,V,t|V_0)$.

\section{Properties of the function $\alpha(W)$}

Starting from its definition~(\ref{alpha}), one can easily obtain the following
properties of the function $\alpha(W)$:
\begin{eqnarray}
&&(i)\quad \alpha(W)\; {\text{is a strictly decreasing function of}}\; W \nonumber\\
&&\\
&&(ii)\quad \alpha(W)=\mu-W\;{\rm for}\; W \leqslant 0\,.\nonumber\\
&& {\text{In particular}},
\alpha(0)=\mu,\; {\text{the first
moment of}}\; \phi(V).\nonumber\\
&&\\
&&(iii)\quad \alpha(W) \rightarrow 0\; {\rm for}\; W\rightarrow \infty.\\
&&(iv)\quad \alpha(W) \rightarrow -\infty\; {\rm for}\; W\rightarrow -\infty.\\
&&(v)\quad \frac{d\alpha(W)}{dW}=-\int_W^{\infty} dU\phi(U).\nonumber\\
&&{\text {Note}}
\left.\frac{d\alpha(W)}{dW}\right|_{W=0}=-1.\\
&&(vi)\quad \frac{d^2 \alpha(W)}{d W^2}=\phi(W)\,.
\end{eqnarray}
Using these relations, one can easily obtain Eq.~(\ref{pdf0}) from Eq.~(\ref{pdf}).

\section{The scaling behavior of $P(V,t)$
for the class I of parent distributions $\phi(V)$}

With the expression~(\ref{alpha1}) of $\alpha^{\rm as}$, one finds from
Eq.~(\ref{pdflongtime}):
\begin{equation}
P(V,t) \approx n_0t {\cal A} V^{\beta}e^{-V^{\delta}}\int_0^V dW\,e^{-e^{S(W)}}\,,
\label{eq5a}
\end{equation}
where
\begin{equation}
S(W)=W^{\delta}-\ln(n_0t{\cal A}/\delta^2)-(\beta+2-2\delta)\ln W\,.
\end{equation}
Consider 
\begin{equation}
W=a_I(t)+y\,,
\end{equation}
where $y$ is small compared to $a_I(t)$. Then
\begin{eqnarray}
&&S(a_I(t)+y)\nonumber\\
&=&[a_I(t)]^{\delta}-\ln\left({n_0t{\cal A}}/{\delta^2}\right)-
(\beta+2-2\delta)
\ln a_I(t)\nonumber\\
&+&\left[\frac{\delta}{[a_I(t)]^{1-\delta}}-\frac{(\beta+2-2\delta)}{a_I(t)}\right]y +{\cal O}(y^2)\,.
\end{eqnarray}
Set
\begin{equation}
[a_I(t)]^{\delta}-\ln(n_0t{\cal A}/\delta^2)-(\beta+2-2\delta)\ln a_I(t)=0\,,
\end{equation}
so, to leading order for large $t$:
\begin{equation}
[a_I(t)]^{\delta} \approx \ln(n_0t{\cal A}/\delta^2)+({\beta+2-2\delta})/{\delta}\,
\ln \ln(n_0t{\cal A}/\delta^2)\,,
\label{eq6}
\end{equation} 
and thus
\begin{equation}
a_I(t) \approx [\ln(n_0t{\cal A}/\delta^2)]^{1/\delta}\,.
\label{eq7}
\end{equation}

Then
\begin{equation}
S(W=a_I(t)+y)=\frac{W-a_I(t)}{b_I(t)}\,,
\end{equation}
where 
\begin{equation}
b_I(t) \approx \frac{1}{\delta} [\ln(nt{\cal A}/\delta^2)]^{(1-\delta)/\delta}\,.
\label{eq8}
\end{equation}

Now, from Eq.~(\ref{eq5a}),
\begin{eqnarray}
&&P(V,t) \approx n_0t b_I(t) {\cal A} V^{\beta} e^{-V^{\delta}}\int_0^VdW\,
e^{-(W-a_I(t))/b_I(t)}\nonumber\\
&&\approx b_I(t) \exp[-V^{\delta}+\ln(n_0t{\cal A})+\beta \ln V]\int_{-\infty}^z 
dU e^{-e^{-U}}\,,\nonumber\\
\label{eq9}
\end{eqnarray}
where $z=(V-a_I(t))/b_I(t)$.
One has
\begin{equation}
\exp[-V^{\delta}+\ln(n_0t{\cal A})+\beta \ln V] \approx \frac{1}{b_I^2(t)}e^{-z}\,,
\label{eq10}
\end{equation}
and thus obtains finally Eq.~(\ref{scaling}) for case I.

\section{The mean square deviation of $n_c(t)$ for $\phi(V)=\exp(-V)$}
\label{msdcol}

Using Eqs.~(\ref{square}) for computing $\langle n_c^2(t)\rangle$ 
For the
particular case of $\phi(V)=\exp(-V)\;(V\geqslant 0)$ one has
$\alpha(V)=\exp(-V)$,
$P(V,t)$ given by Eq.~(\ref{pvexp}),
and the conditional velocity distribution function~(\ref{condvelocity})
\begin{eqnarray}
&&P(V_2,t_2|U,t_1)=e^{-n_0(t_2-t_1)e^{-V_2}}\delta(V_2-U)
\nonumber \\
&&+n_0(t_2-t_1)e^{-V_2}
[Ei(-n_0(t_2-t_1)e^{-U})\nonumber\\
&&-Ei(-n_0(t_2-t_1)e^{-V_2})]\theta(V_2-U)\,.
\end{eqnarray}
Then, after lengthy calculations, Eq.~(\ref{square}) leads to
\begin{eqnarray}
\langle n_c^2(t) \rangle &=&\frac{1}{8}
\ln^2(n_0t)+\left(\frac{C}{2}+\frac{1}{4}\right)\,\ln(n_0t)\nonumber\\
&-&\frac{3}{2 n_0t}-2Ce^{-n_0t}-\frac{e^{-n_0t}}{2n_0t}\nonumber\\
&+&\left(\frac{7}{4}-\frac{C}{2}\right)Ei(-n_0t)-\ln(n_0t)Ei(-n_0t)\nonumber\\
&+&2e^{-n_0t}Ei(-n_0t)+\left(\frac{9C}{4}+\frac{C^2}{2}+\frac{\pi^2}{12}\right)+J\,,\nonumber\\
\end{eqnarray}
where
\begin{eqnarray}
J&=&\frac{1}{2}\,{\mathbf{\cal {P}}}\int_0^{n_0t}\frac{d\tau}{\tau}
\left[e^{-n_0\tau}Ei(-n_0(t-\tau))+Ei(-n_0\tau)\right]\,.\nonumber\\
\end{eqnarray}
Here ${\mathbf{\cal P}}$ designates the principal part of the above integral.

In the long-time limit $n_0t\gg 1$, the main contribution to the value of
$\langle n_c^2(t)\rangle$ comes from the logarithmic terms,
\begin{equation}
\langle n_c^2(t)\rangle \approx \frac{1}{8}
\ln^2(n_0t)+\left(\frac{C}{2}+\frac{1}{4}\right)\,\ln(n_0t)\,.
\end{equation}
Combining this result with the expression~(\ref{ncexp}) of $\langle
n_c(t)\rangle$, one obtains finally
the asymptotic result~(\ref{nc2exp}).

\end{document}